\documentclass[11pt, a4paper, oneside]{Thesis} % Paper size, default font size
\graphicspath{{Pictures/}} % Specifies the directory where pictures are stored
\usepackage[english]{babel}
\usepackage[square, numbers, comma, sort&compress]{natbib} % Use the natbib
\usepackage{hyperref}
\hypersetup{
	colorlinks=true,
	linkcolor=blue,
	filecolor=magenta,      
	urlcolor=cyan,
}
\usepackage{charter}
\usepackage[free-standing-units=true]{siunitx}
\title{\ttitle} % Defines the thesis title - don't touch this

\newcommand{\quotes}[1]{``#1''}
\newcommand{\code}[1]{\texttt{#1}}

\begin{document}

\frontmatter % Use roman numbering style (i, ii...) for the pre-content pages

\setstretch{1.3} % Line spacing of 1.3

% Define page headers using FancyHdr package and set up for one-sided printing
\fancyhead{} % Clears all page headers and footers
\rhead{\thepage} % Sets the right side header to show the page number
\lhead{} % Clears the left side page header

\pagestyle{fancy} % Finally, use the "fancy" page style to implement the
                  %FancyHdr headers

% Input all the variables used in the document. Please fill out the
% variables.tex file with all your details.
%-------------------------------------------------------------------------------
%	DOCUMENT VARIABLES
%
%	Fill in the lines below to set the various variables for the document
%-------------------------------------------------------------------------------

%-------------------------------------------------------------------------------
% Your thesis title - this is used in the title and abstract
% Command: \ttitle
\thesistitle{Disruption Tolerant Networks for Underwater Communications}
%-------------------------------------------------------------------------------
% The document type: Thesis / report, etc.
% Command: \doctype
\documenttype{Bachelors Thesis}
%-------------------------------------------------------------------------------
% Your supervisor's name - this is used in the title page
% Command: \supname
\supervisor{Prof. Mandar \textsc{Chitre}}
%-------------------------------------------------------------------------------
% The supervisor's position - Used on Certificate
% Command: \suppos
\supervisorposition{Associate Professor}
%-------------------------------------------------------------------------------
% Supervisor's institute
% Command: \supinst
\supervisorinstitute{National University of Singapore}
%-------------------------------------------------------------------------------
% Your Co-Supervisor's name
% Command: \cosupname
\cosupervisor{Dr. Suvadip \textsc{Batabyal}}
%-------------------------------------------------------------------------------
% Co-Supervisor's Position - Used on Certificate
% Command: \cosuppos
\cosupervisorposition{Assistant Professor}
%-------------------------------------------------------------------------------
% Co-Supervisor's Institute
% Command: \cosupinst
\cosupervisorinstitute{BITS Pilani, Hyderabad Campus}
%-------------------------------------------------------------------------------
% Your Examiner's name. Not currently used anywhere.
% Command: \examname
\examiner{}
%-------------------------------------------------------------------------------
% Name of your degree
% Command: \degreename
\degree{Bachelor of Engineering (Hons.) Computer Science}
%-------------------------------------------------------------------------------
% The BITS Course Code for which this report is written
% COmmand: \ccode
\coursecode{BITS F421T}
%-------------------------------------------------------------------------------
% The name of the Course
% Command: \cname
\coursename{}
%-------------------------------------------------------------------------------
% Your name. Extend manually in case of multiple authors
% Command: \authornames
\authors{Arnav \textsc{Dhamija}}
%-------------------------------------------------------------------------------
% Your ID Number - used on the Title page and abstract
% Command: \idnum
\IDNumber{\href{mailto:arnav.dhamija@gmail.com}{arnav.dhamija@gmail.com}}
%-------------------------------------------------------------------------------
% Your address
% Command: \addressnames
\addresses{}
%-------------------------------------------------------------------------------
% Your subject area
% Command: \subjectname
\subject{}
%-------------------------------------------------------------------------------
% Keywords for this report.
% Command: \keywordnames
%-------------------------------------------------------------------------------
% University details
% Command: \univname
\university{\texorpdfstring{\href{http://universe.bits-pilani.ac.in/} % URL
                {Birla Institute of Technology and Science}} % University name
                {Birla Institute of Technology and Science}}
%-------------------------------------------------------------------------------
% University details, in Capitals
% Command: \UNIVNAME
\UNIVERSITY{\texorpdfstring{\href{http://universe.bits-pilani.ac.in/} % URL
                {BIRLA INSTITUTE OF TECHNOLOGY AND SCIENCE}} % name in capitals
                {BIRLA INSTITUTE OF TECHNOLOGY AND SCIENCE}}
%-------------------------------------------------------------------------------
% Department Details
% Command: \deptname
\department{\texorpdfstring{\href{http://universe.bits-pilani.ac.in/hyderabad/mechanicalengineering/MechanicalEngineering} % Your department's URL
                {Mechanical Engineering}} % Your department's name
                {Mechanical Engineering}}
%-------------------------------------------------------------------------------
% Department details, in Capitals
% Command: \DEPTNAME
\DEPARTMENT{\texorpdfstring{\href{http://universe.bits-pilani.ac.in/hyderabad/mechanicalengineering/MechanicalEngineering} % Your department's URL
                {MECHANICAL ENGINEERING}} % Your department's name in capitals
                {MECHANICAL ENGINEERING}}
%-------------------------------------------------------------------------------
% Research Group Details
% Command: \groupname
\group{\texorpdfstring{\href{Research Group Web Site URL Here (include http://)}
                {Research Group Name}} % Your research group's name
                {Research Group Name}}
%-------------------------------------------------------------------------------
% Research Group Details, in Capitals
% Command: \GROUPNAME
\GROUP{\texorpdfstring{\href{Research Group Web Site URL Here (include http://)}
                {RESEARCH GROUP NAME (IN BLOCK CAPITALS)}}
                {RESEARCH GROUP NAME (IN BLOCK CAPITALS)}}
%-------------------------------------------------------------------------------
% Faculty details
% Command: \facname
\faculty{\texorpdfstring{\href{Faculty Web Site URL Here (include http://)}
                {Faculty Name}}
                {Faculty Name}}
%-------------------------------------------------------------------------------
% Faculty details, in Capitals
% Command: \FACNAME
\FACULTY{\texorpdfstring{\href{Faculty Web Site URL Here (include http://)}
                {FACULTY NAME (IN BLOCK CAPITALS)}}
                {FACULTY NAME (IN BLOCK CAPITALS)}}
%-------------------------------------------------------------------------------

%-------------------------------------------------------------------------------
%   NON-CONTENT PAGES
%-------------------------------------------------------------------------------
\begingroup
\hypersetup{hidelinks}
\maketitle
% \Declaration
%\Certificate

\begin{abstract}
Disruption Tolerant Networks (DTNs) are employed in applications where the network is likely to be disrupted due to environmental conditions or where the network topology makes it impossible to find a direct route from the sender to the receiver. Underwater networks typically use acoustic waves for transmitting data. However, these waves are susceptible to interference from sources of noise such as the wake from ships, sounds from snapping shrimp, and collisions from acoustic waves generated by other nodes.

DTNs are good candidates for situations where successfully delivering the message is more important than low delivery times and high network throughput. This is true for certain applications of underwater networks. DTNs can also create new options for network topologies, such as opening up the possibility of using \quotes{data muling} nodes if the network is resilient to delays.

The Acoustic Research Laboratory (ARL) at NUS has developed their own Groovy-based underwater network simulator called \emph{UnetStack}, in which network protocols can be designed and tested in a simulator. These protocols can later be directly deployed on physical hardware, such as Subnero's underwater modems. Hence, this project revolves around creating a new UnetStack protocol called \code{DtnLink} for enabling disruption tolerant networking in various use cases of the ARL.
\end{abstract} 

%\begin{acknowledgements}
%I would like to thank Prof. Mandar Chitre, NUS for helping and motivating me for this project. The weekly meetings helped me maintain momentum and Prof. Mandar's insights were invaluable in helping me sort out any technical issues I had while giving me autonomy for the implementation. I would also like to thank Dr. Suvadip Batabyal, BITS Hyderabad for co-supervising me and advising me on the evaluation components for the same.
%\end{acknowledgements}

%-------------------------------------------------------------------------------
%	LIST OF CONTENTS/FIGURES/TABLES PAGES
%-------------------------------------------------------------------------------

% The page style headers have been "empty" all this time, now use the "fancy"
% headers as defined before to bring them back
\pagestyle{fancy}
	\lhead{\emph{Contents}}
	\tableofcontents
	\lhead{\emph{List of Figures}}
	\listoffigures
	\listoftables
\endgroup

%-------------------------------------------------------------------------------
%	ABBREVIATIONS
%-------------------------------------------------------------------------------

\clearpage % Start a new page

 % Set the line spacing to 1.5, this makes the following tables easier to read
\setstretch{1.5}

\lhead{\emph{Abbreviations}} % Set the left side page header to "Abbreviations"
\listofsymbols{ll} % Include a list of Abbreviations (a table of two columns)
{
\textbf{ARL} & \textbf{A}coustic \textbf{R}esearch \textbf{L}aboratory \\
\textbf{AUV} & \textbf{A}utonomous \textbf{U}nderwater \textbf{V}ehicle \\
\textbf{DTN} & \textbf{D}isruption \textbf{T}olerant \textbf{N}etwork \\
\textbf{EM} & \textbf{E}lectro\textbf{m}agnetic \\
\textbf{JVM} & \textbf{J}ava \textbf{V}irtual \textbf{M}achine \\
\textbf{MTU} & \textbf{M}aximum \textbf{T}ransmission \textbf{U}nit \\
\textbf{PDU} & \textbf{P}rotocol \textbf{D}ata \textbf{U}nit \\
\textbf{RFC} & \textbf{R}equest \textbf{F}or \textbf{C}omments \\
\textbf{SCAF} & \textbf{S}tore \textbf{C}arry \textbf{A}nd \textbf{F}orward \\
\textbf{TTL} & \textbf{T}ime \textbf{T}o \textbf{L}ive \\
}

%-------------------------------------------------------------------------------
%	PHYSICAL CONSTANTS/OTHER DEFINITIONS
%-------------------------------------------------------------------------------
%
% FIXME: a definitions page might be good here, maybe some footnotes too

%-------------------------------------------------------------------------------
%	SYMBOLS
%-------------------------------------------------------------------------------

\clearpage % Start a new page

%-------------------------------------------------------------------------------
%	DEDICATION
%-------------------------------------------------------------------------------

\setstretch{1.25} % Return the line spacing back to 1.3

%\addtocontents{toc}{\vspace{2em}} % Add a gap in the Contents, for aesthetics

%-------------------------------------------------------------------------------
%	THESIS CONTENT - CHAPTERS
%-------------------------------------------------------------------------------

\mainmatter % Begin numeric (1,2,3...) page numbering

\pagestyle{fancy} % Return the page headers back to the "fancy" style

% Include the chapters of the thesis as separate files from the Chapters folder
% Uncomment the lines as you write the chapters

% Chapter 1

\chapter{Introduction}

\label{Chapter1}

\lhead{Chapter 1. \emph{Introduction}}

%----------------------------------------------------------------------------------------

\section{Overview}
\subsection{Disruption Tolerant Networks}

Disruption Tolerant Networks (DTNs) are used in a number of applications where conventional communication schemes are inadequate due to erratic network conditions, lack of network infrastructure, or long propagation delays in the communication medium. Unlike conventional network protocols which rely on end-to-end connectivity at a given instant of time, DTNs do \emph{not} require a complete path from the source to the destination when transmitting the message.

Furthermore, some DTN protocols \citep{Spyropoulos2010} create multiple copies of the messages, expecting at least one of these copies to opportunistically reach the destination node. All types of DTN protocols employ a type of Store-Carry-And-Forward (SCAF) mechanism to store the message until it can be sent to the destination. The message can either be sent directly to the destination or via another node in multi-hop routing.

This makes DTNs very useful for sending data when the network used inherently unreliable due to environmental conditions and when delivery is prioritised over network throughput. For example, NASA used their own implementation of a DTN to communicate with the ISS from Earth \citep{Marshall2010}.

\subsection{Underwater Acoustic Networks}

Underwater wireless communication is a developing field \cite{underwater_issues}, which presents several issues which are not typically encountered in terrestrial wireless networks. For one, electromagnetic waves do not propagate through water due its high dielectric constant, so conventional RF wireless protocols can not be used. Instead, acoustic waves are used for encoding and transmitting information. However, being based on sound waves, this is much more susceptible to interference from sources of noise such as the wake from ships, sounds from animals, and collisions from acoustic waves generated by other nodes. In particular, Singapore is a challenging environment for deploying underwater acoustic networks due to the noise created from its busy shipping industry. Singapore also is the natural habitat of snapping shrimp, which produce a distinct sound wave which interferes with these acoustic waves \cite{Legg2012}. Due to all these issues, there is a higher probability of transmitted messages being dropped due to the lossy channel medium than there is typical RF networks. Hence, the network is more likely to be \emph{disrupted}.

Acoustic waves travel at the speed of sound in water (around \SI{1500}{m/s}), which is several orders of magnitude slower than EM waves which travel at the speed of light. This can result in high propagation delays ($d_{prop}$). The bitrates of acoustic networks is low, usually in the order of 5 KB/sec. This leads to longer transmission delays ($d_{trans}$). Processing of acoustic waves can involve significant error correction and signal processing which contributes to a higher processing delay as well ($d_{proc}$). Putting all this together, we get the following delay for sending a single message:

\[ d_{end-to-end} = d_{prop} + d_{trans} + d_{proc}\]

Therefore, delays can be significant in underwater networks and protocols need to be designed which take these delays into account.

We can see that underwater networks can be affected by both \emph{disruptions} and \emph{delays} in the channel medium. DTN protocols are meant to alleviate the affect of both of these issues, making them a good fit for underwater networks.

\subsubsection{UnetStack}

The \emph{Unet} (Underwater Networks) project is jointly developed by the Acoustics Research Laboratory (ARL) and its commercial partner, Subnero\footnote{\href{https://subnero.com/}{https://subnero.com/}}. UnetStack \citep{Chitre2015} is an agent-based network simulator which is used for testing the protocols that are used in real-world deployments of underwater networks. UnetStack uses a \emph{software-in-the-loop} network stack based on the Java Virtual Machine (JVM) which allows protocols developed in UnetStack to be directly deployed on hardware. It has APIs for Groovy, Java, Python, and C.

\begin{figure}[h!]
	\centering
	\includegraphics[width=0.6\linewidth]{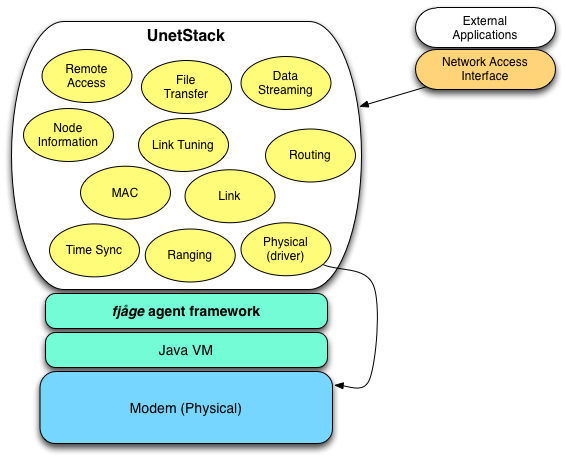}
	\caption{UnetStack Agent Architecture}
	\label{unetstack}
\end{figure}

UnetStack does not use the conventional layered network model. Instead, the network stack is divided into \quotes{agents} which handle different concerns of the network. For example, the \code{ROUTING} agent handles the management of routes of messages and the \code{PHYSICAL} agent can be used as a driver for an underwater modem. Messages can be directly passed from one agent to another. This flexibility is particularly important in underwater networks where network bandwidth is at a premium.

\section{Use Cases} \label{usecase}

This project is about developing a new \code{LINK} agent which will implement a DTN protocol. This agent will be called \code{DtnLink} throughout this report. It is designed for the use cases of some of ARL's projects. 

Some of these are as follows:

\begin{itemize}
	\item \textbf{Data Muling}: UnetStack is used on sensor nodes for collecting sensor measurements from parts of the ocean. The sensor stores the data until a diver can retrieve the sensor. This is a labour intensive procedure. To supplant this, \code{DtnLink} can be used for sending the sensor's data to an AUV \cite{Chitre2008} when it comes in range of the sensor. Unet AUV's have sophisticated algorithms for navigating towards a sensor for establishing a link for communication \citep{Doniec2013}.
	\item \textbf{Time Varying Links}: A concern in underwater networks is that certain links are only available under certain conditions. For example, high bandwidth optical links are short-ranged and require a Line of Sight to the destination for communication. Ideally, \code{DtnLink} should be able to choose the most optimal link depending on the link's availability and bitrate.
	\item \textbf{USB Link}: \code{DtnLink} will maintain a list of pending messages in the node's non-volatile storage. As a potential alternative to sending these messages wirelessly, a USB Link agent could work in conjunction with \code{DtnLink} for automatically copying these messages to an external storage device.
	\begin{figure}[h!]
		\centering
		\includegraphics[scale=0.6]{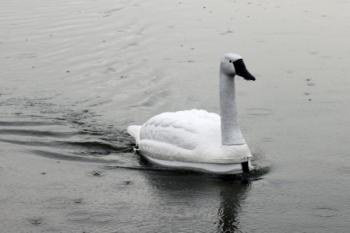}
		\caption{The NUSwan Robot}
		\label{swan}
	\end{figure}
	\item \textbf{NUSwan}: The \emph{NUSwan} \citep{Koay2015} in Figure \ref{swan} is a water surface dwelling robot which autonomously collects data about the water quality in Singapore's reservoirs with its sensors. This data is relayed to the cloud using an LTE connection. However, in large reservoirs, the LTE connection may be temporarily unavailable due to lack of coverage. \code{DtnLink} can store pending messages and then send these messages when the LTE link is available.
\end{itemize}

\section{Modelling a DTN Protocol for Underwater Networks}

\code{DtnLink} aims to be a drop-in addition to UnetStack for adding disruption tolerant communication support. Hence, it is essential to define some of the features which are required for disruption tolerant communication in underwater networks:

\subsection{Node Advertisement Messages} \label{beacon}
As previously mentioned, underwater communication is adversely affected by packet collisions. Hence, a pending message should only be sent when the sender is within communication range of another node to avoid flooding the network with messages which cannot reach the destination. To accomplish this, \code{DtnLink} SHOULD periodically send a message without any data at a set interval to advertise its existence to nearby nodes (a so-called \quotes{Beacon} message). On receiving this Beacon message, a node can start sending datagrams to the Beacon's sender.

UnetStack nodes also have the capability to \emph{snoop} on the messages destined for other nodes sharing the same physical medium for communication. This capability is used for discovery of other nodes without having to send an explicit Beacon message. Additionally, \code{DtnLink} SHOULD NOT send an additional Beacon message if it has sent a datagram on a particular link in that time period.

\code{DtnLink} MUST support the capability to store datagrams on the non-volatile storage of nodes until it can be sent to the destination. It MUST also delete datagrams whose TTL has expired.

Note that the working of this Beacon functionality is under the assumption that the connectivity of the links is symmetric. That is, if Node A can receive a transmission from Node B, Node B is also able to receive a transmission from Node A. However, this assumption may not be valid for certain underwater applications.

\subsection{Required Features}\label{features}

\begin{itemize}
	\item \textbf{Storage}: \code{DtnLink} MUST store datagrams on the node's non-volatile storage until the datagram can be sent to the destination.
	\item \textbf{TTL}: Datagrams saved to the node's non-volatile storage MUST be deleted when the TTL of the datagram is exceeded. TTL information for a datagram MUST be propagated through the network. \code{DtnLink} SHOULD do so by encapsulating a datagram in its own PDU as described in Section \ref{pdu}. As each node may not have its clock in sync with other nodes, TTL SHOULD be stored as the time left till the message expires instead of an expiry time for a particular node.
	\item \textbf{Reliability}: \code{DtnLink} only uses Link agents supporting reliability for sending messages. Hence, we are guaranteed to know if a datagram has failed or has been successfully delivered. \code{DtnLink} SHOULD forward a \code{DatagramDeliveryNtf} to the requesting application. On the other hand, if a datagram times out, \code{DtnLink} SHOULD send a \code{DatagramFailureNtf} to the application.
	\item \textbf{Node Advertisement}: As explained in Section \ref{beacon}, \code{DtnLink} SHOULD periodically send \quotes{Beacons} on all its underlying Link agents for alerting other nodes about its presence. On receiving a Beacon, a node can start sending messages residing in its non-volatile storage to that node.
	\item \textbf{Multiple Links}: A particular node may have multiple available Link agents. \code{DtnLink} SHOULD populate a list of all the Link agents which support reliability. \code{DtnLink} MAY also automatically switching between links depending on whether they have a connection to the next hop for a message.
	\item \textbf{Power Failure Recovery}: Power failure on a node will typically cause the node to drop all pending messages in its buffers. \code{DtnLink} MAY implement a mechanism of gracefully recovering from power loss by resending messages which are pending in the node's non-volatile storage provided their TTLs have not expired.
	\item \textbf{Fragmentation}: Messages which exceed the MTU of the underlying links MAY be split by \code{DtnLink} into smaller fragments which are sent like regular message. If the implementation supports fragmentation, the receiving instance of \code{DtnLink} MUST wait for the reception of all of these fragments before reassembling the original message. Fragments MUST be encoded in the \code{DtnLink's} PDU format.
	\item \textbf{Randomised Sending}: While a very rare issue in real-world deployments, message sent at exactly the same time can result in collisions in simulations. Hence, \code{DtnLink} MAY delay sending message by a random amount of time.
	\item \textbf{Stop-And-Wait Sending}: To avoid congesting the channel medium, \code{DtnLink} MAY adopt the strategy of only sending one message at a time and waiting for a notification about its receipt before sending the next one.
	\item \textbf{Short-circuit Sending}: As implemented by the newer UnetStack3 agents, \code{DtnLink} MAY support short-circuiting messages on single-hop routes by sending the message without its PDU headers. This reduces the message size. Regardless, messages exceeding the MTU will still need to be encoded in PDUs to be reassembled at the receiver's instance of \code{DtnLink}. However, messages sent through short-circuiting may be duplicated at the receiver.
	\item \textbf{Single-copy Routing}: Some DTN routing algorithms use packet replication to send messages to the destination. This approach might be sub-optimal for underwater networks which are constrained by transmission power limitations and suffer from packet collisions when the network is flooded with messages. Therefore, an implementation of DTNs for UnetStack MAY NOT use packet replication.
\end{itemize}

From this set of requirements, we can identify which ones can be included in our protocol. The \code{DtnLink} agent supports all of theses features, including the optional ones as illustrated in Chapter \ref{Chapter2}.

% Chapter 2

\chapter{Design}

\label{Chapter2}

\lhead{Chapter 2. \emph{Design}}

\code{DtnLink} is written in Apache Groovy, the lingua franca of the \emph{fjåge} and the \emph{Unet} project \cite{8286399}. Groovy runs on the JVM and can be used either statically and dynamically. This allows it to be fully compatible with all Java code and its associated libraries.

\section{Message Sending}

\subsection{The DtnLink PDU}\label{dtnpdu}
\begin{figure}[h!]
	\centering
	\includegraphics[width=0.6\linewidth]{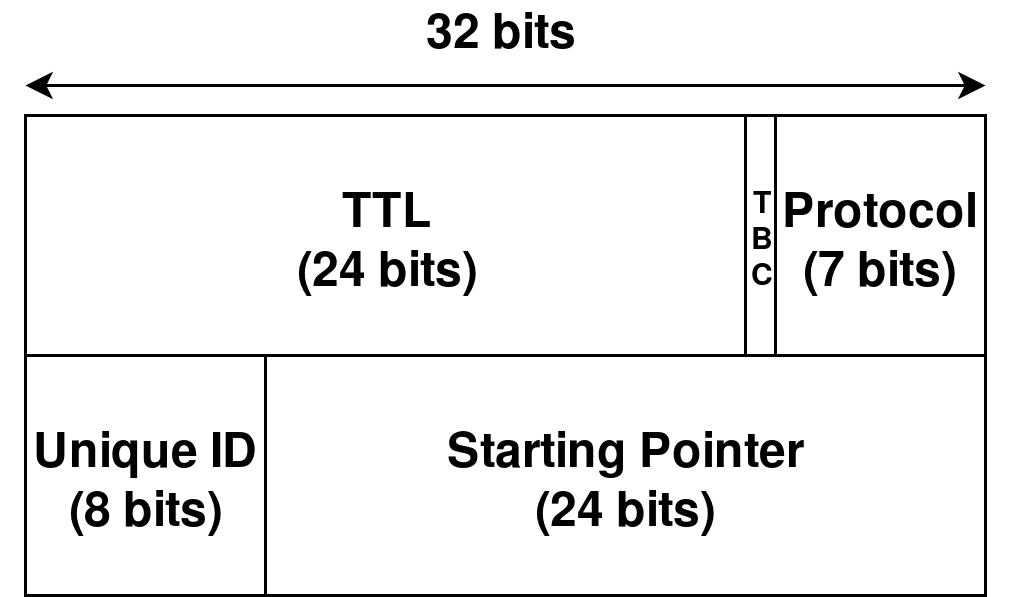}
	\caption{The DtnLink Protocol Data Unit (PDU)}
	\label{pdu}
\end{figure}

Before sending messages, \code{DtnLink} encodes the data in a PDU (Protocol Data Unit) which is encapsulated in a \code{DatagramReq} before sending on a link. 
\newpage
The structure of this PDU is shown in Figure \ref{pdu}. The following is the data represented in this PDU:

\begin{itemize}
	\item 24-bit TTL, representing the lifetime of the message in seconds.
	\item 1-bit To Be Continued (TBC) bit, for informing the receiver if more fragments are expected for large messages which do not fit in the \code{LINK}'s MTU. A value of 0 indicates the transmission is complete for that payload.
	\item 7-bit Protocol number of the original message. This is used by UnetAgents for identifying which \code{DatagramNtfs} are intended for them.
	\item 8-bit Unique ID, for uniquely identifying messages and distinguishing payloads by the tuple of their sender and the Payload ID.
	\item 24-bit Starting pointer, for informing the receiver about where to insert the contents of a fragment into its payload file.
\end{itemize}

A \code{DatagramReq} is the data structure used for sending messages between agents in UnetStack. This PDU is generated when the \code{DtnLink} receives a \code{DatagramReq} containing the the message from another agent. Before sending to the destination node, the message's TTL is updated. Therefore the \code{DtnLink} PDU helps in tracking the message's TTL, identifying duplicate messages, and managing the sending of large messages (\emph{payloads}).

The following examples illustrate the different cases handled by the \code{DtnLink} agent.

\subsection{Single-Hop Message Delivery}\label{subsec:singlehop}

In these examples, we can see how the \code{DtnLink} sends messages to the destination by encoding the information in its PDU format, described in Section \ref{dtnpdu}. 

In the below figures, a UnetAgent application (App/1) on Node 1 wants to send a message via its \code{DtnLink} agent (DTNL/1). \code{DtnLink} encodes the message in its PDU and then it uses an underlying \code{ReliableLink} (RL/1). The blue part of the figure indicates the message being transmitted physically underwater. After reception of the message at the modem of Node 2, the ReliableLink (RL/2) will pass the message upto the node's \code{DtnLink} (DTNL/2). Here, the message will be decoded, the message information will be extracted and passed onto the application of Node 2 (App/2).

\begin{figure}[!h]
	\centering
	\includegraphics[width=0.6\linewidth]{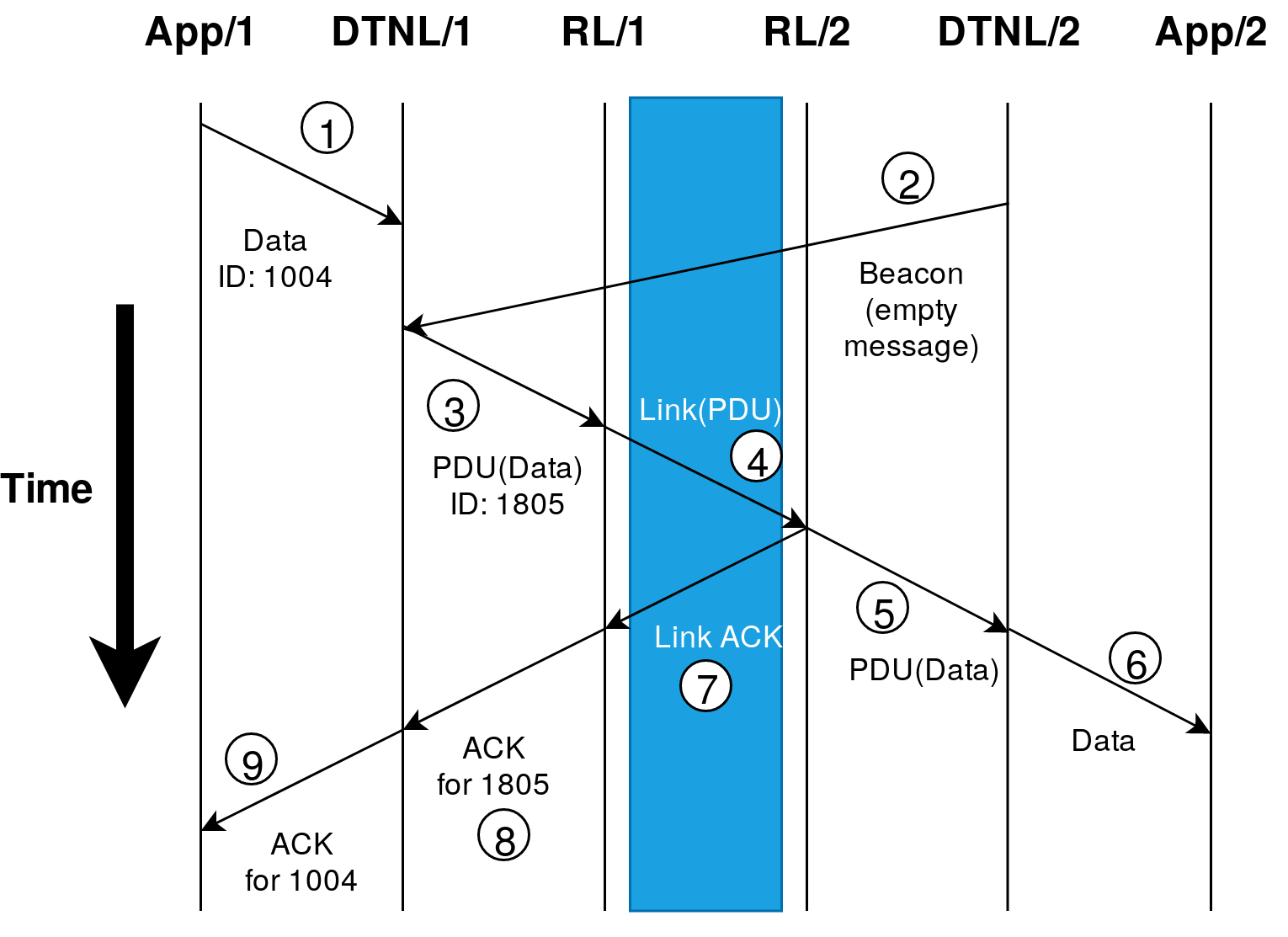}
	\caption{Single-Hop Delivery}
	\label{shd}
\end{figure}

In Figure \ref{shd} we can see how the \code{DtnLink} sends messages when the destination is the next hop in the network. In this case, \code{DtnLink} waits until the destination node is online by receiving its Beacon message and then sends the datagram. On successful delivery acknowledgement from the underlying link, the ACK, called a \code{DatagramDeliveryNtf} in UnetStack terminology, is passed up to the application.

\begin{figure}[h!]
	\centering
	\includegraphics[width=0.6\linewidth]{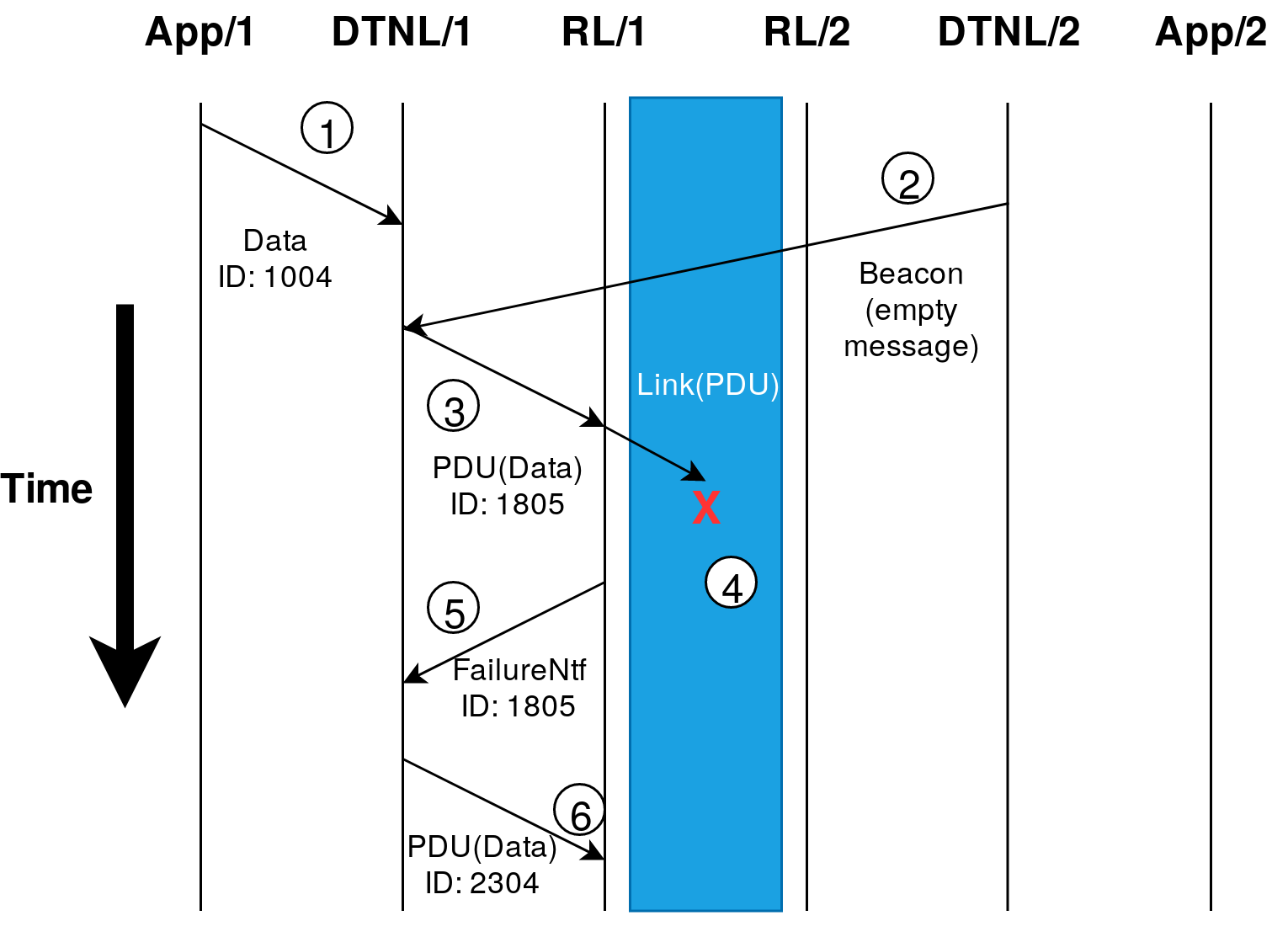}
	\caption{Single-Hop Failure}
	\label{shf}
\end{figure}

Figure \ref{shf} illustrates an example of failure of sending a datagram. In case if the sender does not receive a \code{DatagramDeliveryNtf} (ACK) before its timeout period, the underlying link on the sender will send a \code{DatagramFailureNtf} which is received by the \code{DtnLink}. As failing to send a message at a particular point of time is \emph{not} necessarily failure in DTNs, the \code{DtnLink} will attempt to send the message at a later point of time when the destination node is online.

\begin{figure}[h!]
	\centering
	\includegraphics[width=0.6\linewidth]{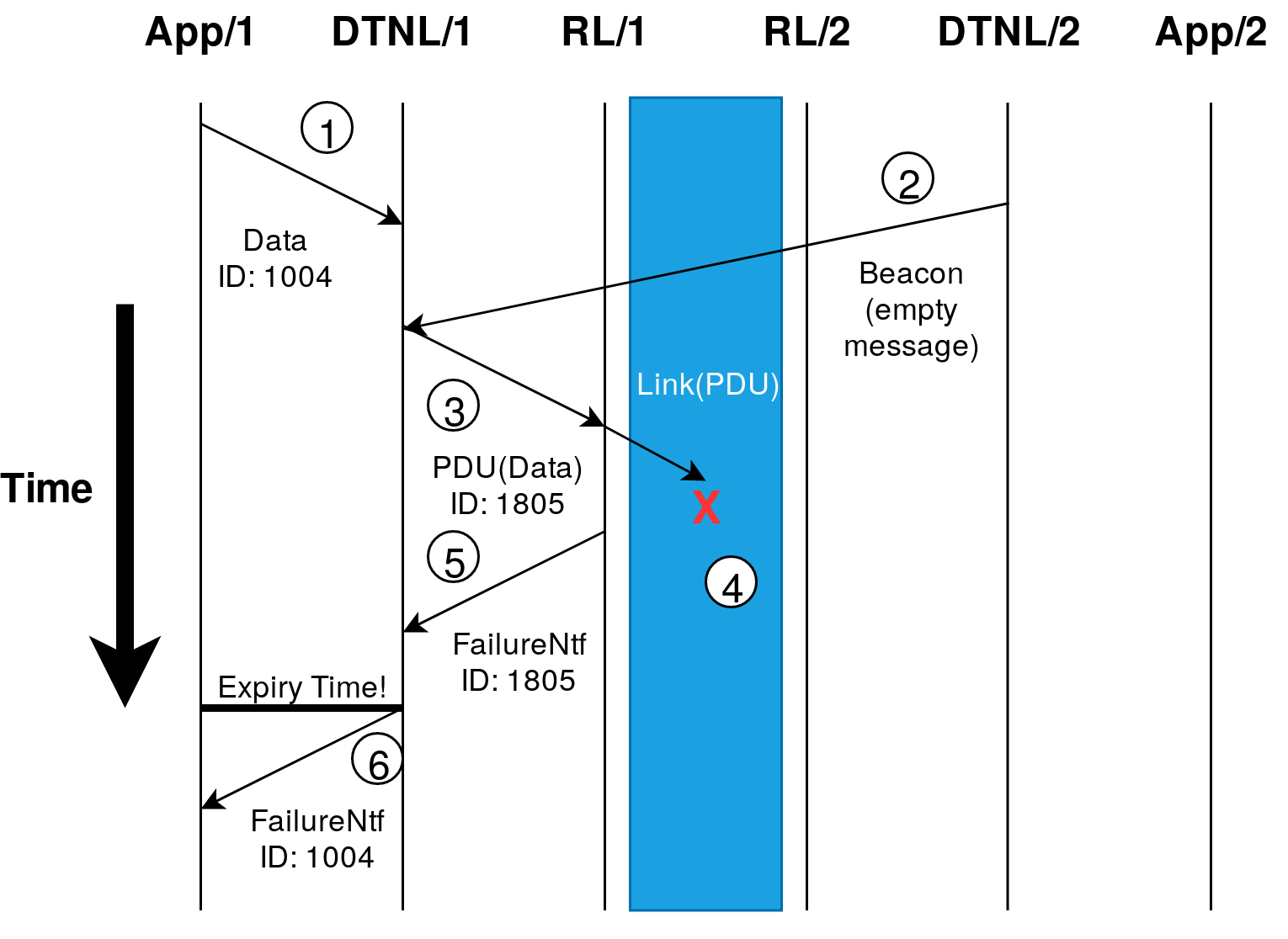}
	\caption{Single-Hop TTL Expiry}
	\label{sht}
\end{figure}

TTL expiry as shown in Figure \ref{sht} can occur when the destination node is not online during the lifetime of the message. In these cases, the message is deleted from the sender and it can no longer be sent in any circumstances. The \code{DtnLink} informs the requesting application about this failure with a \code{DatagramFailureNtf}.

\begin{figure}[h!]
	\centering
	\includegraphics[width=0.6\linewidth]{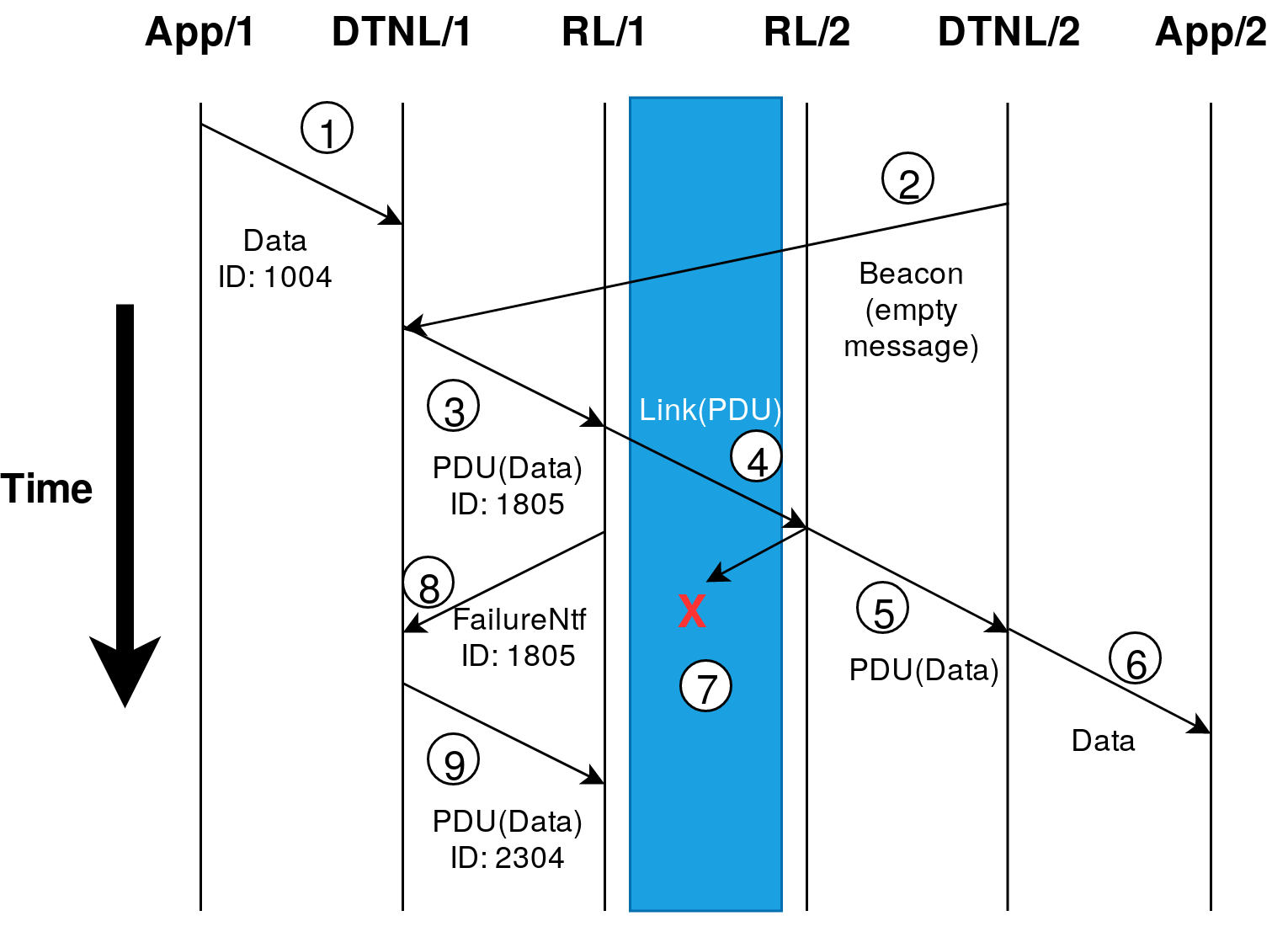}
	\caption{Single-Hop ACK Failure}
	\label{shaf}
\end{figure}

In the final case, Figure \ref{shaf} shows a problematic scenario in which the \code{DatagramDeliveryNtf} (ACK) message is lost due to a lossy channel medium. When this happens, the link on the sender's side will timeout and will generate a \code{DatagramFailureNtf} for \code{DtnLink}. This will make the \code{DtnLink} resend the message, resulting in the receiver receiving duplicate messages. Clearly, this problem must be avoided by having \code{DtnLink} check for duplicate messages.

\subsection{Duplicate Message Detection}\label{dupcheck}

As shown in Figure \ref{shaf}, a dropped ACK can cause a message to be sent repeatedly until a \code{LINK} level ACK is received. This can cause duplicate messages at the receiver.

\code{DtnLink} solves this by encoding a random, 8-bit Unique ID in the PDU (Section \ref{dtnpdu}) for each message. When a receiver receives a message, it computes the \code{hashCode} of the entire message after excluding the TTL field of the PDU. This value is stored in a \code{Set} in the receiver's instance of \code{DtnLink}. If the generated \code{hashCode} does not exist in this \code{Set}, the message is sent to the application, else the message is discarded.

\subsection{Short Circuit Message Sending}\label{sec:sc}

\begin{figure}[!h]
	\centering
	\includegraphics[width=0.6\linewidth]{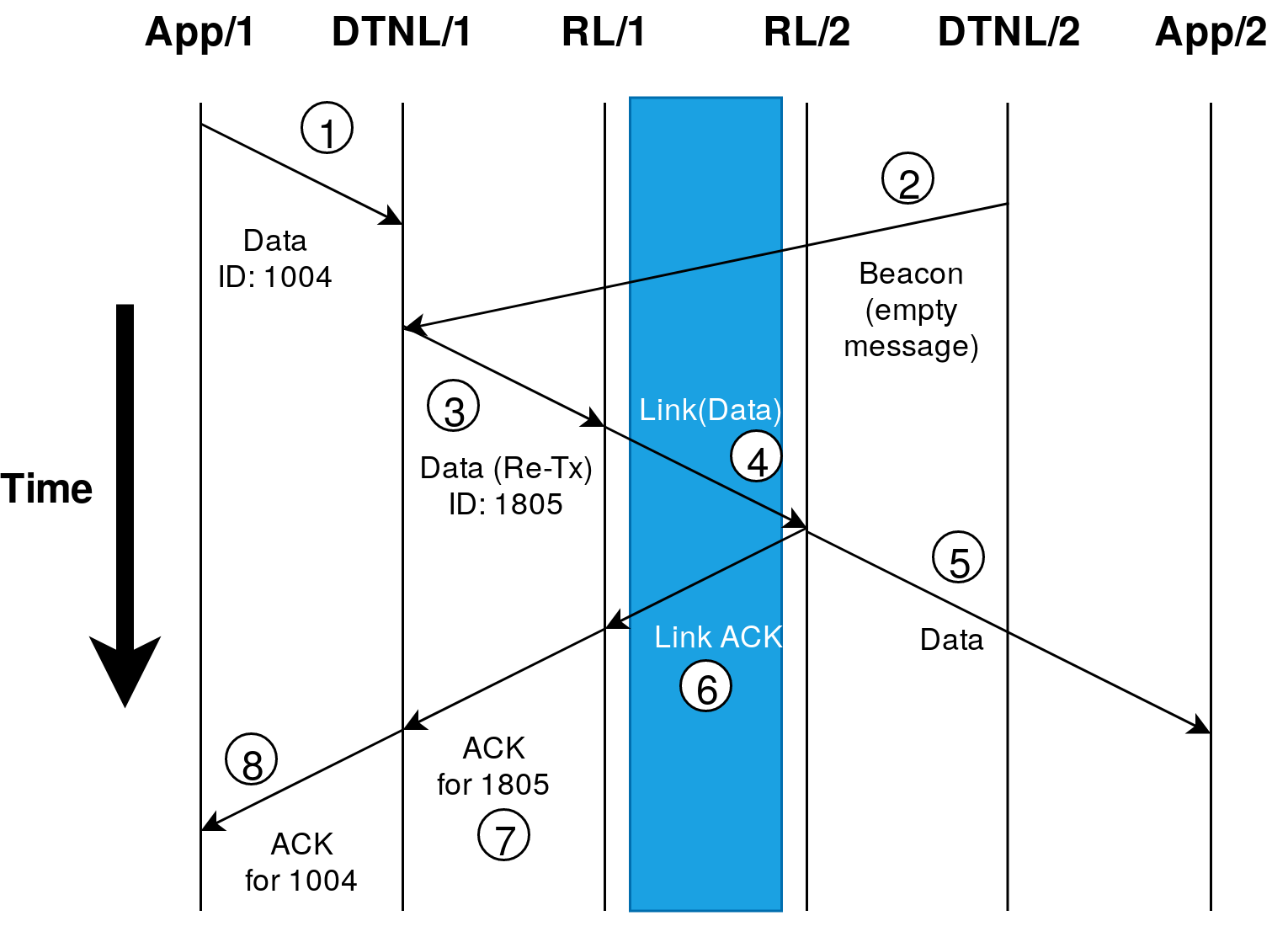}
	\caption{Short Circuiting Messages}
	\label{shortcircuitfig}
\end{figure}

In some cases, we might choose to send the message directly without encoding the \code{DtnLink} PDU headers to reduce message size. In this case, we can \emph{short circuit} the message, as briefly mentioned in Section \ref{features}. Figure \ref{shortcircuitfig} shows how a message is transmitted straight to the desired agent without encoding it in the \code{DtnLink} PDU format.

However, the trade-off of short circuiting is that it can only work for single-hop messages and it eschews the duplication message detection mechanism which was explained in Section \ref{dupcheck}.

\section{Power Failure Recovery}

Disruption tolerant networks can also be affected by disruptions in the network infrastructure. For instance, it's possible that a battery powered node runs out of charge in the middle of a mission or a solar powered buoy loses power on a cloudy day. Ideally, we would want our protocol to be able to \emph{gracefully} recover from such disruptions and keep the pending messages intact for sending in the future.

\code{DtnLink} is capable of recovering from the situation in which a node is unexpectedly shutdown and the \code{DtnLink} agent is terminated. This is implemented by saving the next hop of a message and its expiry time along with its PDU on the node's non-volatile storage. On startup \code{DtnLink} scans its directory for pending messages which have not yet expired. The next hop and expiry time of a message is used in rebuilding the \code{metadataMap} which is used by \code{DtnStorage} for tracking pending messages. Once this is done, \code{DtnLink} can send the messages via the strategies discussed in Section \ref{subsec:singlehop}.

\section{Components}

\code{DtnLink} has been designed with modularity in mind. The following classes were created to implement \code{DtnLink}.

\begin{itemize}
	\item \code{DtnLink}: \code{DtnLink} extends \code{UnetAgent} and handles the sending and receiving of messages. As explained in Section \ref{beacon} and Section \ref{dupcheck} it also sends Beacon messages and checks for duplicate messages. It also responds to \code{DatagramDeliveryNtf} and \code{DatagramFailureNtf} messages and sends messages according to the priority set by the user.
	
	\code{DtnLink} is configured to only send datagrams on \code{LINK} agents which support the \code{RELIABILITY} capability. Supporting \code{RELIABILITY} does not mean that the \code{LINK} will always be able to successfully send the message. Instead, it means that the \code{LINK} agent is able to generate acknowledgements for every message sent.

	When the \code{DtnLink} finds a new node (either through a probe or a snooped message), it will query this data structure for the PDUs destined for the node. Once this is done, the TTLs are checked for expiry and sent by a \code{LINK} agent.
	
	As we are exclusively using \code{LINK} agents with \code{RELIABILITY} we are \emph{guaranteed} to get a acknowledgement about the result of the delivery. If \code{DtnLink} is notified of a successful transmission, the entry is deleted from the tracking \code{Hashmap} in \code{DtnStorage} and the corresponding PDU file is deleted along with it. If the \code{DtnLink} receives a notification about delivery failure, it attempts to send the message at a later time when the node is within the transmission range.
	
	\item \code{DtnStorage}: This will handle the storage mechanism. It will track outbound PDUs, fragment and reassemble payload messages, and will delete expired PDUs. Expired messages are deleted periodically at an interval which can be set by the user. It also encodes and decodes into a format which can be used by \code{DtnLink}. It can also restore the state of \code{DtnLink} after restarting from power failure.

	\item \code{DtnLinkManager}: \code{DtnLink} is expected to handle a variety of \code{LINK} agents. A node can have a number of communication media, such as an underwater acoustic modem, optical link, and Ethernet tether. Some nodes may only support one of these communication media. The \code{DtnLinkManager} class maintains data structures which store information about the properties of each \code{LINK} agent that a node supports. It also maintains lists of the links supported by the node's neighbouring nodes. These data structures are updated on every message received by the node. Furthermore, a user can also set the priority of links used for communicating between two nodes which share more than one common \code{LINK}.
	
	\item \code{DtnPduMetadata}: This class allows \code{DtnStorage} to track the messages which it has sent to other nodes. Each message ID has a corresponding \code{DtnPduMetadata} object which records the next hop destination of the message, its expiry time, and the number of bytes of the message successfully transmitted for payloads.
\end{itemize}

\section{Capabilities}

This agent will support the \code{LINK} and \code{DATAGRAM} service. Other agents can forward messages with a valid TTL value to the \code{DtnLink} for disruption tolerant delivery. Messages without a valid TTL will be refused outright.

In its current iteration, \code{DtnLink} only support single-copy and single-hop routing. If required, \code{ROUTING} agents can be used in conjunction with \code{DtnLink} for multi-hop purposes.

\section{Configurable Options}{\label{sec:config}}
\code{DtnLink} is highly configurable and the following Parameters can be adjusted depending to the use-case:
\begin{itemize}
	\item \textbf{Short circuit}: As explained in Section \ref{sec:sc}, short circuiting messages can reduce message size. However, this parameter is turned off by default as short circuiting messages makes it impossible to check for duplicate receptions of a particular message.
	\item \textbf{Periodic Functions}: The \code{beaconTimeout} (maximum time of the link being idle before sending a Beacon message), \code{GCPeriod} (time period of deleting expired and delivered datagrams from non-volatile storage), \code{datagramResetPeriod} (time period of sending datagrams), and \code{linkExpiryTime} (time for which a link can remain idle without removing it from the active links list) parameters can be set at runtime.
	\item \textbf{Datagram Priority}: Messages can be sent according to their order of \code{ARRIVAL}, ascending order of \code{EXPIRY} times, and in a \code{RANDOM} manner. These options are exposed in the \code{datagramPriority} parameter.
	\item \textbf{Link Priority}: The order in which underlying links are used by \code{DtnLink} can be changed by sending a list of the AgentIDs to \code{linkPriority}. If these AgentIDs are \code{null} or not registered by the \code{DtnLink}, the request will be ignored.
\end{itemize}

\subsection{Automated Regression Testing}
\code{DtnLink} can be tested reproducibly. As new features are added to the agent, it is imperative that a basic subset of its functionality remains intact. These tests check that the key features of \code{DtnLink} are working correctly.

As the \code{DtnLink} will work in conjunction with several other agents, it is more useful to see the output of the agent on certain inputs rather than diving into the  implementation of how each function performs. This is formally called ``Black-box'' testing.

\begin{figure}[h!]
	\centering
	\includegraphics[width=0.9\linewidth]{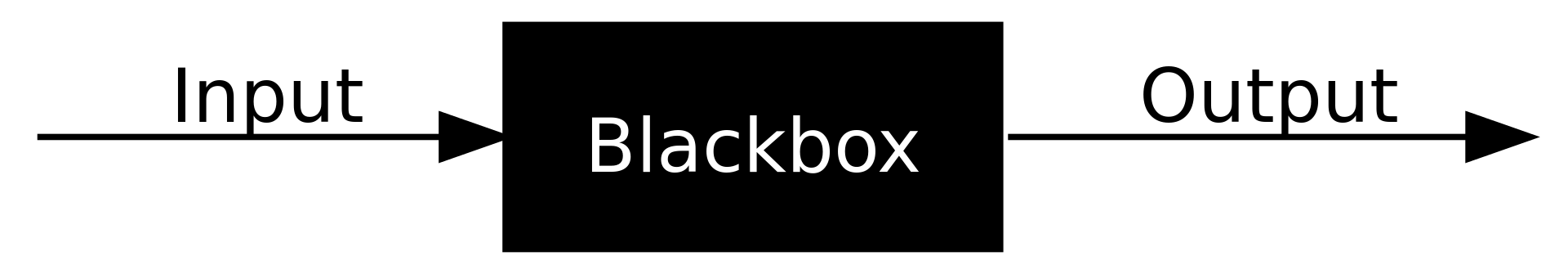}
	\caption{Black-box testing}
	\label{bb}
\end{figure}

The above figure is a simple example of the key concept of the black-box. The internals can be totally abstracted for the tests as we only wish to see the outputs of the black-box on certain inputs. In these tests, the \code{DtnLink} is the black-box and the specially developed \code{TestApp} and \code{TestLink} agents test the behaviour of the \code{DtnLink}. More specifically, the \code{TestApp} prepares \code{DatagramReqs} for sending to the \code{DtnLink} and the \code{TestLink} checks the receipt of these datagrams, and send the corresponding \code{Ntfs} to the \code{DtnLink}.

\begin{figure}[h!]
	\centering
	\includegraphics[width=0.6\linewidth]{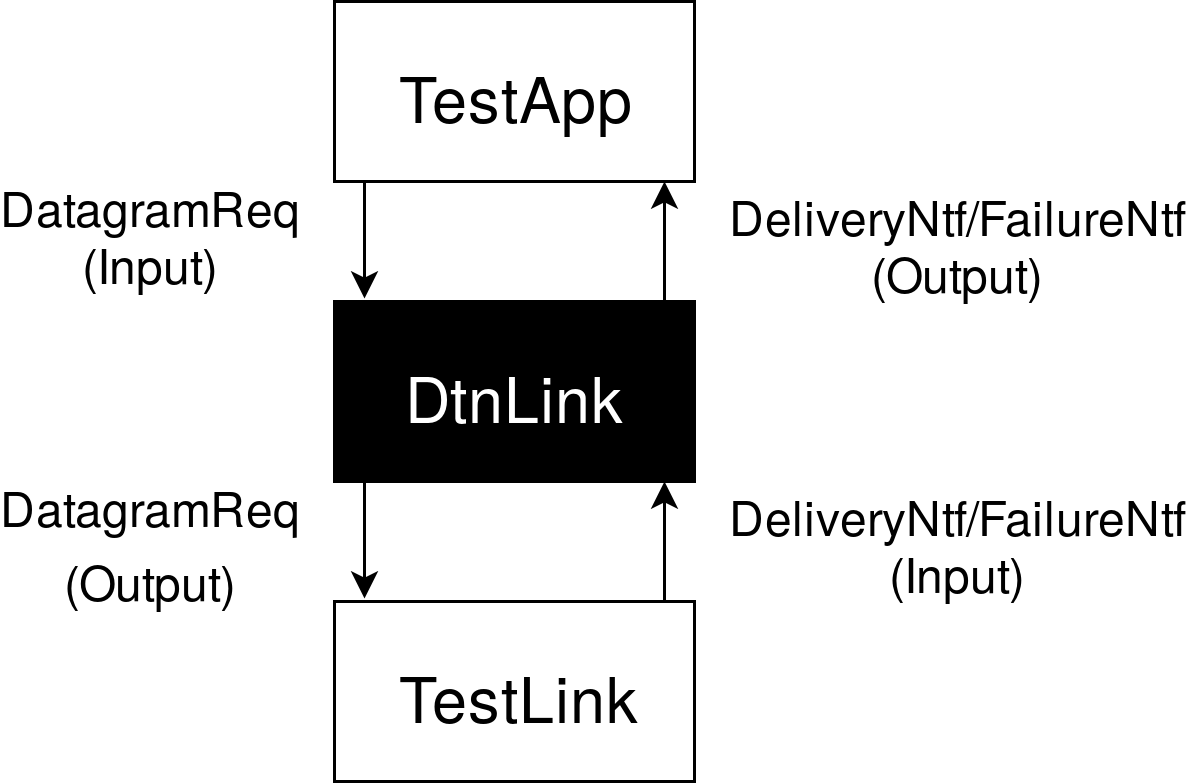}
	\caption{The DtnLink Test Harness}
	\label{dtnharness}
\end{figure}

Figure \ref{dtnharness} shows what the test harness of \code{DtnLink} looks like. In this, we have created a new \code{TestLink} and \code{TestApp} class which sends \code{DatagramReq} messages (input) to \code{DtnLink}. We can then check if \code{DtnLink} produces the correct messages (output) at the App and Link parts of the test.

By these means, we can ``trick'' the \code{DtnLink} into behaving as it would in a multi node simulation. The following tests are conducted with this test suite, using \code{JUnit}\footnote{\href{https://junit.org/junit5/}{https://junit.org}}:

\begin{itemize}
	\item \code{TRIVIAL\_MESSAGE}: This test sends an empty \code{DatagramReq} to \code{DtnLink} to check if the agent correctly accepts messages with TTLs encoded.
	\item \code{SUCCESSFUL\_DELIVERY}: This test sends a \code{DatagramReq} with the \code{USER} protocol number to check if the message sent to the underlying link is sent without the \code{DtnLink} headers and can be short circuited. It also checks whether the \code{DatagramReq} is formatted correctly and has the original Protocol number.
	\item \code{ROUTER\_MESSAGE}: This test sends a \code{DatagramReq} with the \code{ROUTING} protocol number to check if the message sent to the underlying link is encoded correctly with the \code{DtnLink} PDU scheme and has its TTL adjusted accordingly.
	\item \code{BAD\_MESSAGE}: This test checks if the \code{DtnLink} responds with a \code{Performative.REFUSE} when it receives a \code{DatagramReq} without a set TTL value.
	\item \code{EXPIRY\_PRIORITY}: This test checks if the messages sent to \code{DtnLink} in \code{EXPIRY\_PRIORITY} mode from \code{TestApp} are forwarded to the \code{TestLink} in order ascending order of their TTL values.
	\item \code{ARRIVAL\_PRIORITY}: This test checks if the messages sent to \code{DtnLink} in \code{ARRIVAL\_PRIORITY} mode from \code{TestApp} are forwarded to the \code{TestLink} in order ascending order of their arrival times.
	\item \code{RANDOM\_PRIORITY}: This test checks if the messages sent to \code{DtnLink} in \code{RANDOM\_PRIORITY} mode from \code{TestApp} are forwarded to the \code{TestLink} in random order without regards to the TTL values or arrival time.
	\item \code{LINK\_TIMEOUT}: This test checks if \code{DtnLink} correctly disables sending messages on links which have not sent a message for a certain period of time.
	\item \code{MULTI\_LINK}: This test checks if \code{DtnLink} correctly uses the priority of Links set through a \code{ParameterReq} to change the priority of the links used to communicate with other nodes.
	\item \code{PAYLOAD\_MESSAGE}: This test checks if the \code{DtnLink} is capable of correctly fragmenting a large message into smaller fragments to fit in the underlying link's MTU. These fragments are sent to another instance of \code{DtnLink} to check if they fragments can be successfully reassembled to form the original datagram.
	\item \code{REBOOT}: This test simulates the behaviour of \code{DtnLink} in event of a power failure. It runs two instances of \code{DtnLink}, one after the other. In the first instance, the test sends messages to \code{DtnLink} for storage and fails all its attempts to transmit the message successfully. The result of this is that the \code{DtnLink}'s directory will be populated with unsent messages. After this, another instance of \code{DtnLink} is created. This test checks that \code{DtnLink} successfully rebuilds its \code{metadataMap} and transmits the messages residing in its internal storage.
\end{itemize}
% Chapter 3

\chapter{Simulations \& Results}

\label{Chapter3}

\lhead{Chapter 3. \emph{Simulations \& Results}}

The \code{DtnLink} agent aims to improve the reliability of sending messages for real-world applications of UnetStack. To better understand how well underwater network protocols work, UnetStack includes a simulator in which underwater nodes running Unet protocols can be simulated. As the communication media is often lossy, UnetStack supports multiple underwater communication models such as the Protocol Channel Model, Basic Acoustic Model, Mission 2013a Model \cite{Chitre2015}, and Urick Acoustic Model \citep{urick1983principles}. 

In the Protocol Channel Model, we can adjust the values of \code{pDetection} which allows us to simulate varying levels of disruption. \code{pDetection} is the probability a node will be able to detect a signal which is within the node's detection range. Lossy channels have a low value of \code{pDetection}.

By simulating various scenarios with \code{DtnLink}, we can better understand how messages can be sent in a disruption tolerant manner.
\newpage
\section {Scenarios}

\subsection{DTN Multihop}\label{sec:dtn_hops}

\begin{figure}[h!]
	\centering
	\includegraphics[width=\linewidth]{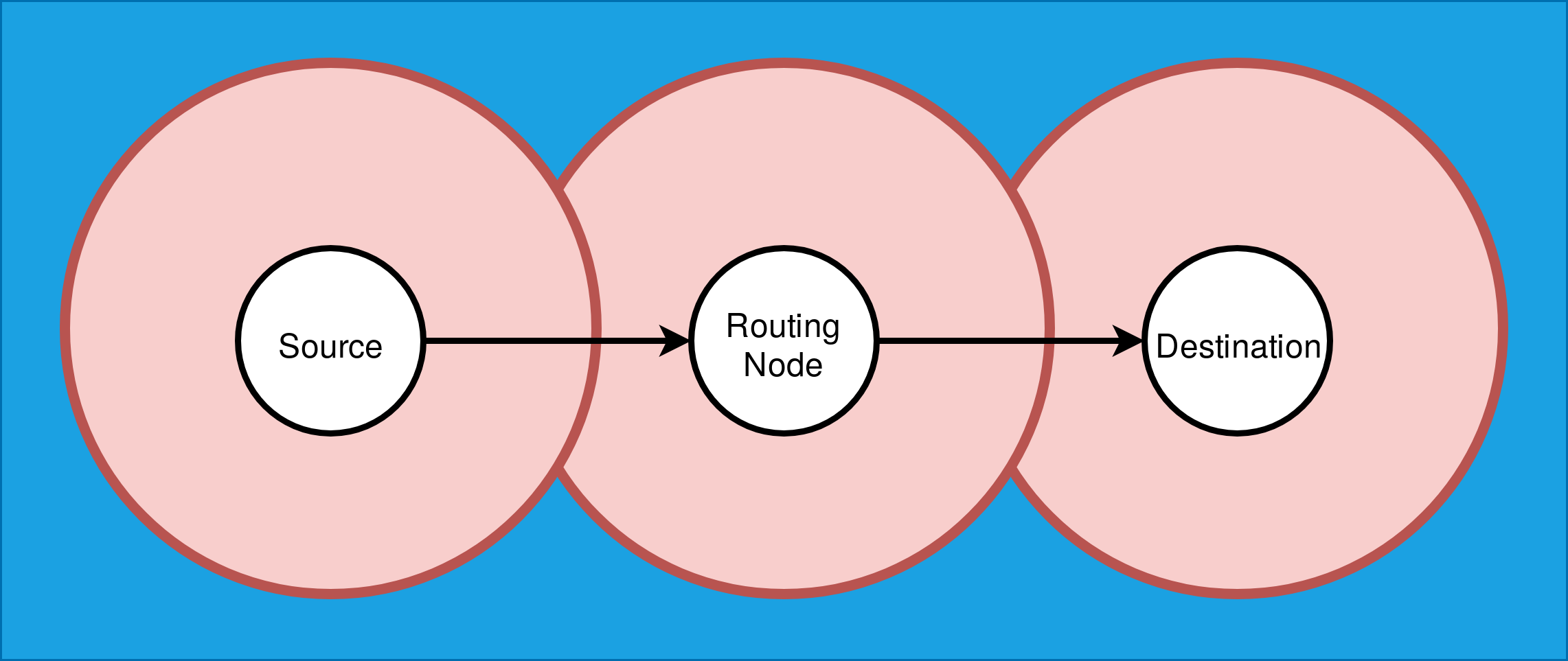}
	\caption{Multihop scenario}
	\label{multihop}
\end{figure}

In this scenario (illustrated in Figure \ref{multihop}), we have a set of three nodes which are placed in such a way that the sender and receiver are out each other's communication range (shaded in red). Hence, in this scenario, we require an intermediate node which relay messages. In this simulation, we can compare the performance of \code{DtnLink} to the typically used \code{ReliableLink} to see how effective it is in lossy networks. The details of the parameters used in this simulation are given in Table \ref{tab:multihopparam} and Table \ref{tab:multihopnodes}.

\begin{table}[!h]
	\centering
%	\resizebox{0.5\linewidth}{!}{%
		\begin{tabular}{|l|l|}
			\hline
			\textbf{Parameter} & \textbf{Value} \\
			\hline
			Simulation Time & \SI{10800}{\second} \\
			Communication Range & \SI{1500}{\meter} \\
			Message Size & \SI{40}{bytes} \\ 
			Message Frequency & \SI{10}{\second} \\
			Message TTL & \SI{10800}{\second} \\
			Total Messages sent from Source & 200 \\
			\hline
		\end{tabular}
%	}
	\caption{Multihop simulation parameters}
	\label{tab:multihopparam}
\end{table}

\begin{table}[!h]
	\centering
	%	\resizebox{0.5\linewidth}{!}{%
	\begin{tabular}{|c|c|c|c|}
		\hline
		\textbf{Node} & \textbf{X} & \textbf{Y} & \textbf{Z} \\
		\hline
		Source & \SI{0}{\metre} & \SI{0}{\metre} & \SI{-50}{\metre} \\
		Routing Node & \SI{1500}{\metre} & \SI{0}{\metre} & \SI{-50}{\metre} \\
		Destination & \SI{3000}{\metre} & \SI{0}{\metre} & \SI{-50}{\metre} \\
		\hline
	\end{tabular}
	\caption{Node co-ordinates for the multihop scenario}
	\label{tab:multihopnodes}
\end{table}

The entire simulation is run for different values of \code{pDetection} for both \code{DtnLink} and \code{ReliableLink}.

\newpage
\begin{figure}[h!]
	\centering
	\includegraphics[width=\linewidth]{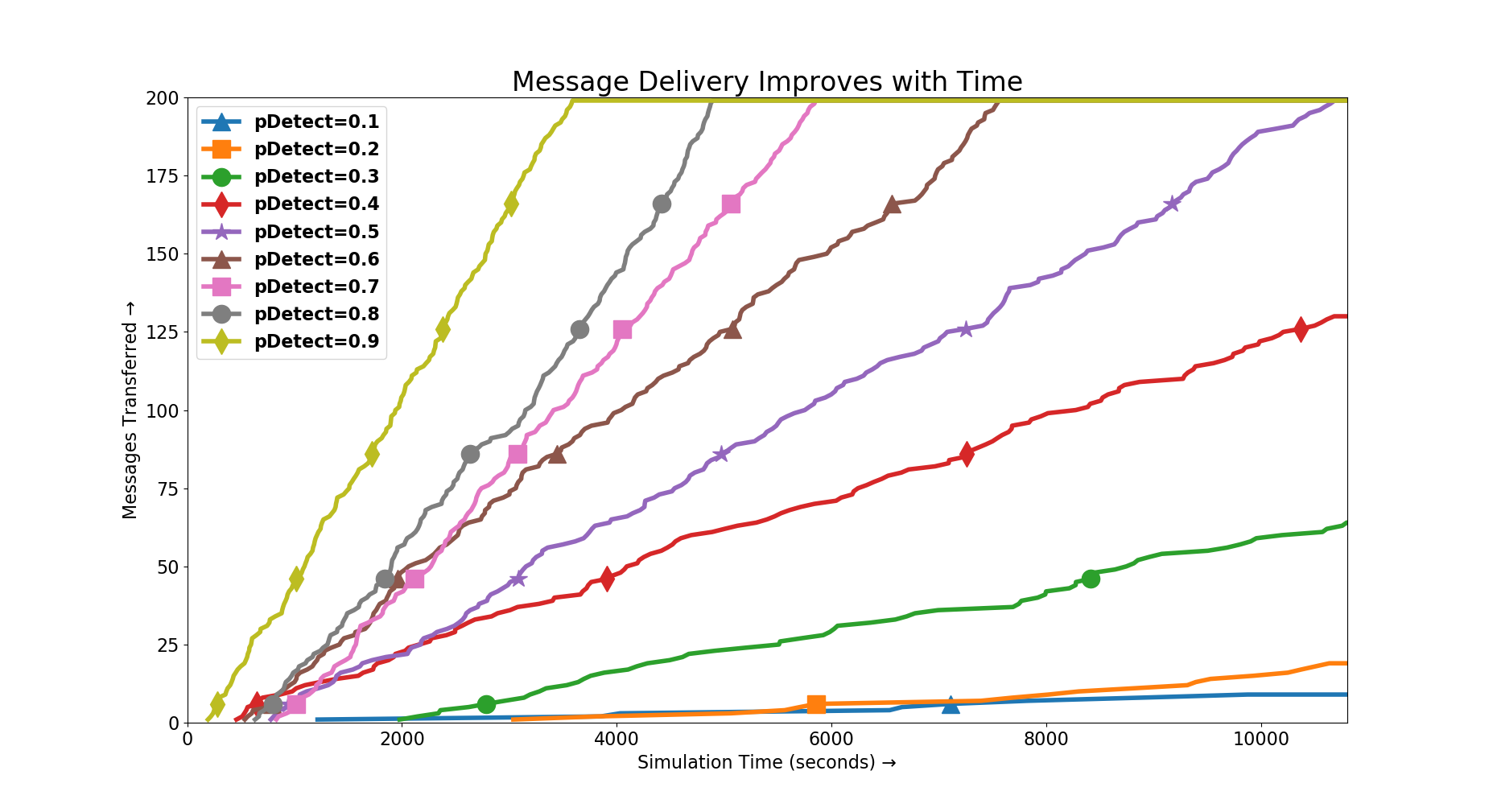}
	\caption{Effect of pDetection on message delivery versus time}
	\label{9plots}
\end{figure}

In Figure \ref{9plots}, we can see that the number of messages delivered increases significantly as the simulation time increases. At lower levels of \code{pDetection} (0.1--0.4), the number of messages transferred is much lower as the reception of several messages fail due to the lossy channel medium. However, given that the simulation runs for enough time and the message's TTL does not expire, \code{DtnLink} will be able to eventually transfer all the messages sent by the sender.

When \code{pDetection} varies from 0.5--0.9, we can see that all 200 messages sent by the sender reach the destination node within the time frame of this simulation. Here, we can see that the time taken to transfer all the messages is affected by \code{pDetection}. This is because a message needs to be retried more times when \code{pDetection} is low.
\newpage

For seeing if \code{DtnLink} is beneficial when we are using lossy networks with a low value of \code{pDetection}, it is useful if we can compare it to the performance of a commonly used \code{LINK} agent which does not have explicit support for disruption tolerance. In the following figures, we can see how \code{DtnLink} compares\footnote{\code{DtnLink} uses \code{ReliableLink} as the underlying link for actually transmitting messages over the channel medium. Hence, it can be expected that at the very least, that \code{DtnLink} should not perform worse than \code{ReliableLink}. Nevertheless, these simulations are illustrative of in which scenarios using \code{ReliableLink} as the underlying agent of \code{DtnLink} can be beneficial compared to using it without \code{DtnLink}} with using the popular \code{ReliableLink} agent for sending messages.

\begin{figure}[h!]
	\centering
	\includegraphics[width=\linewidth]{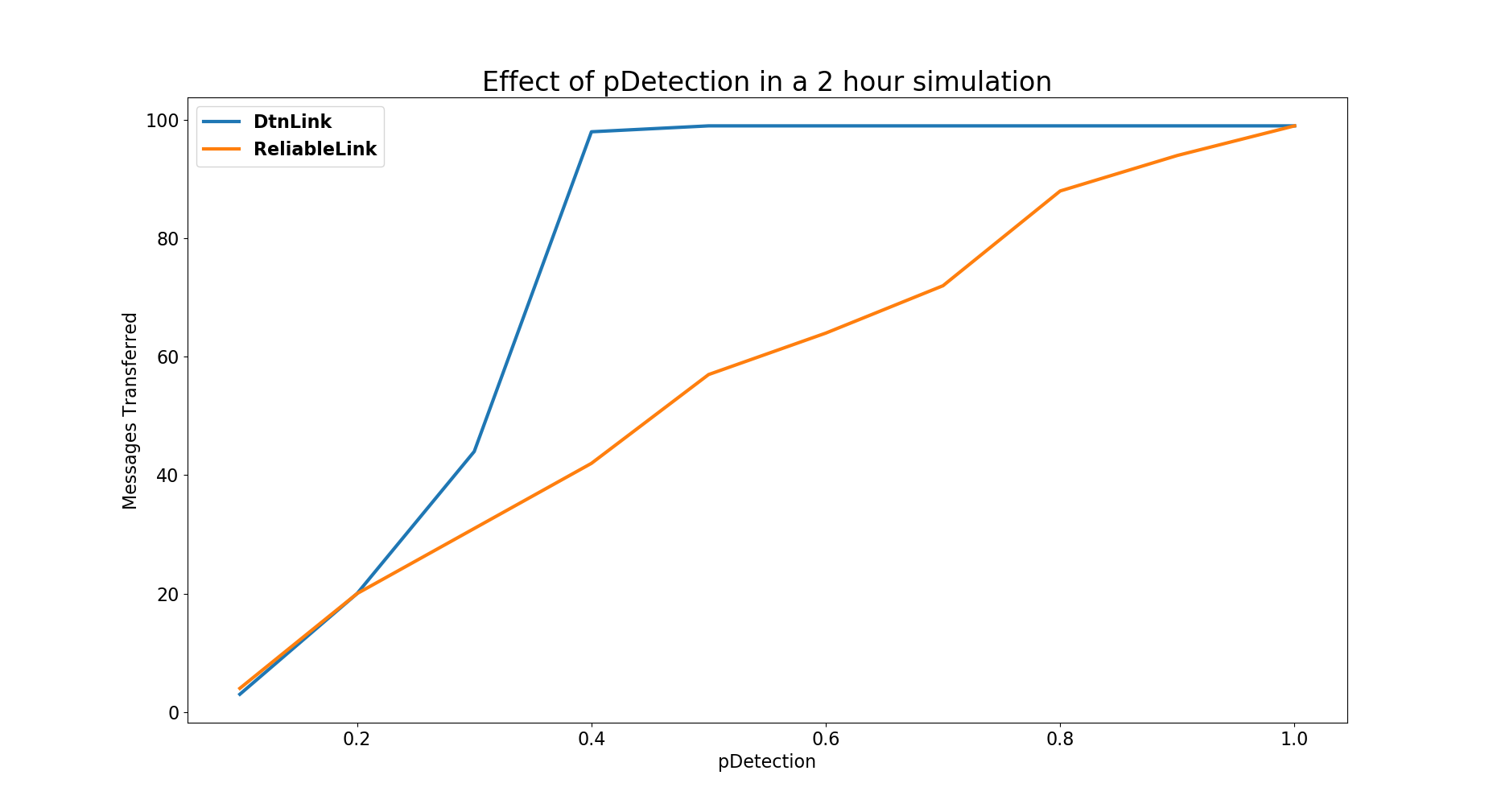}
	\caption{Comparison of ReliableLink and DtnLink for different values of pDetection}
	\label{2hour}
\end{figure}

In Figure \ref{2hour}, we can see that for a set of 100 messages, \code{DtnLink} outperforms \code{ReliableLink} over a simulation time of \SI{7200}{\second}. This is due to \code{DtnLink}'s capability of being able to retry the message until it is successfully delivered. \code{ReliableLink}'s probability of successfully delivering the message is directly a function of \code{pDetection} as it will only retry a failed message until its \code{maxRetries} (default = 3) limit is exceeded.

\begin{figure}[h!]
	\centering
	\includegraphics[width=\linewidth]{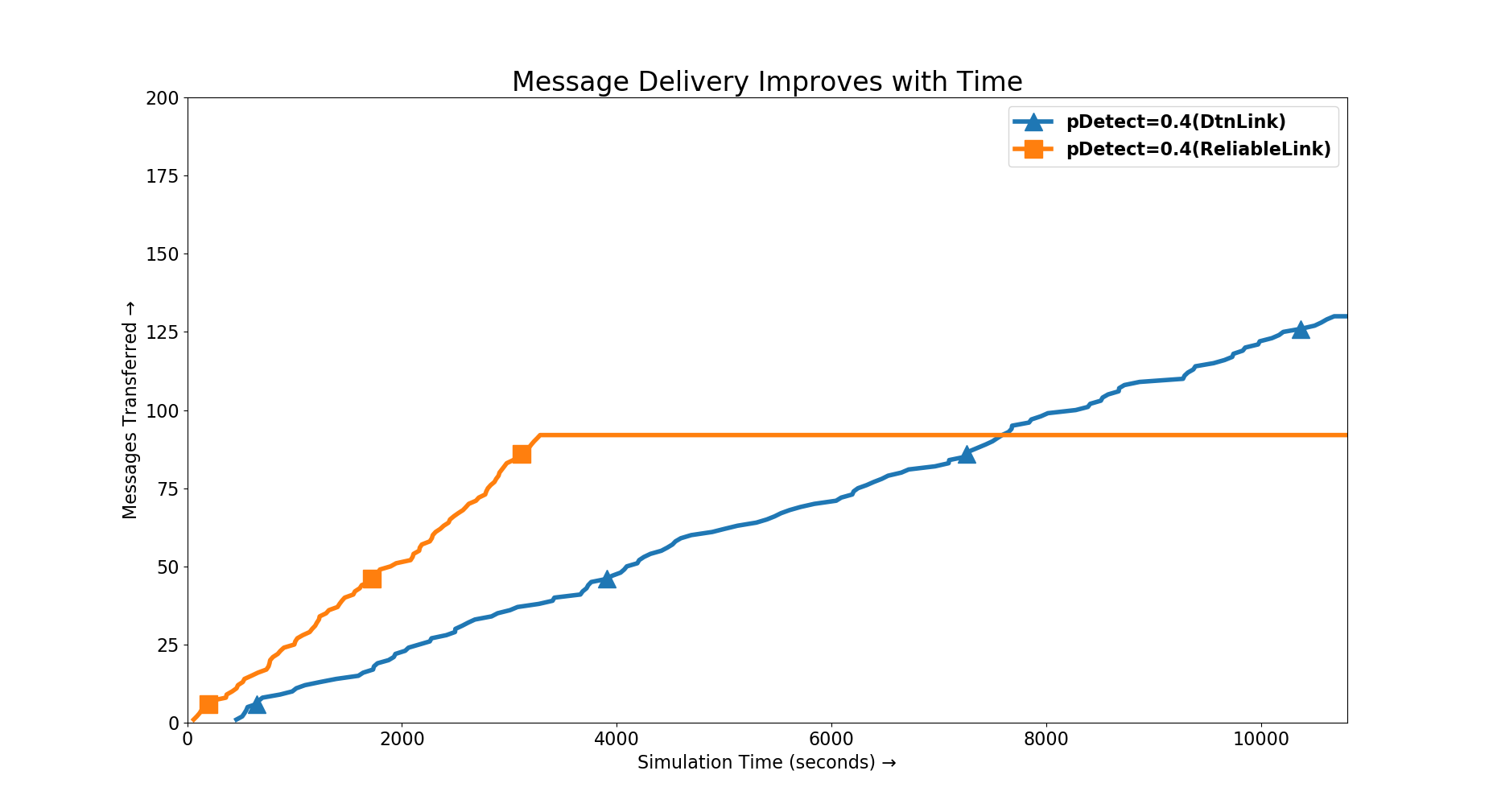}
	\caption{Comparison of ReliableLink and DtnLink for fixed value of pDetection}
	\label{dvrl4}
\end{figure}

However, at \code{pDetection} from 0.1--0.2, we can see that \code{DtnLink} does not offer much advantage over \code{ReliableLink}. The reason behind this is \code{DtnLink}'s Stop-And-Wait protocol of sending messages which limits the number of messages it can send in a given amount of time. If the channel is very lossy, \code{DtnLink} will spend a long amount of time waiting for the result of a message being delivered. However, as shown in Figure \ref{9plots}, the message will be transferred given more time. This effect is more apparent in Figure \ref{dvrl4}.
\newpage

In Figure \ref{dvrl4}, we can see that the messages delivered by \code{DtnLink} only exceeds that of \code{ReliableLink} after a certain amount of time. This shows that the Stop-And-Wait sending method of \code{DtnLink} can negatively impact delivery times. Hence, \code{DtnLink} is more useful when used in an application which can tolerate long delays.

\subsection{AUV Data Muling}\label{sec:auv_mule}

\begin{figure}[!h]
	\centering
	\includegraphics[width=0.9\linewidth]{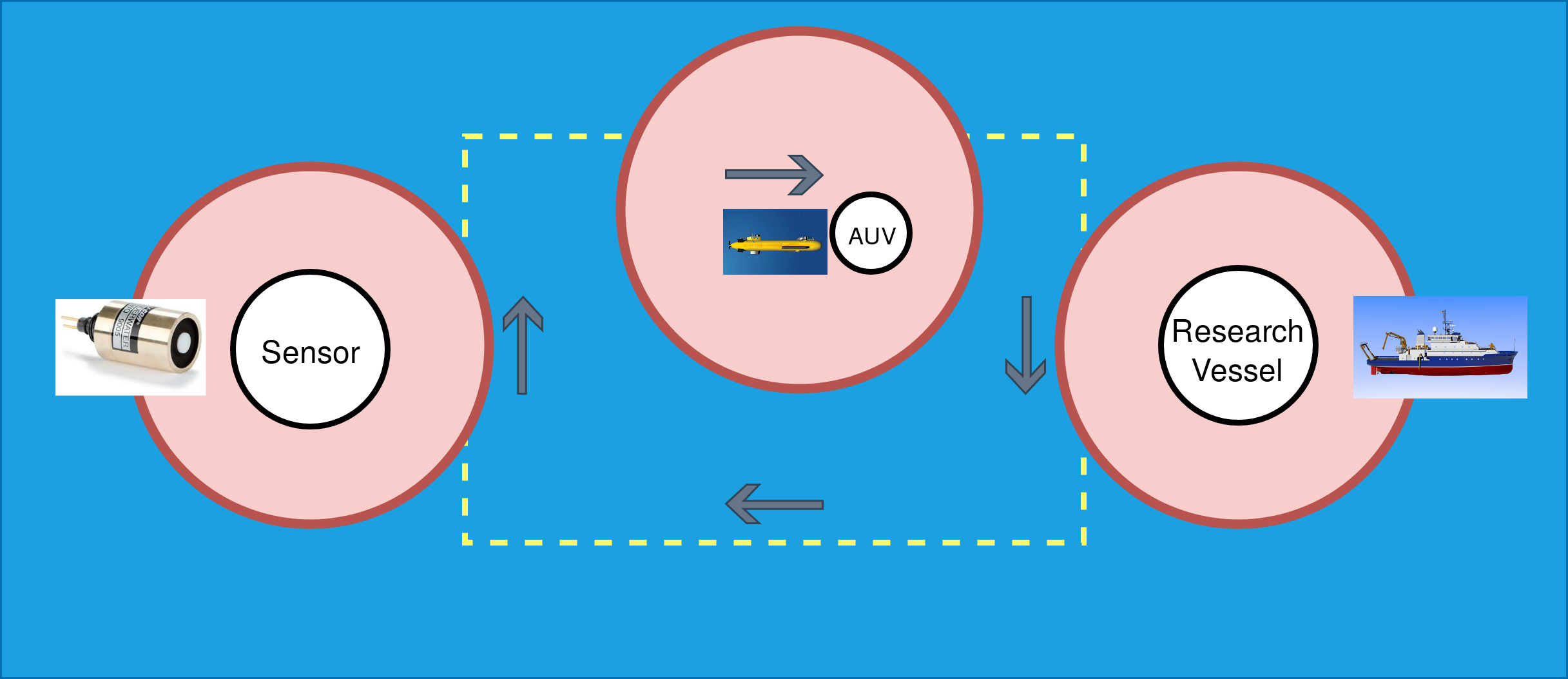}
	\caption{AUV Data Muling scenario}
	\label{fig:mulecomm}
\end{figure}

In Figure \ref{fig:mulecomm} we can see an example of an AUV carrying messages between two nodes - that is the sensor and the research vessel. This use case, described in Section \ref{usecase} is a variant of the multihop demonstration shown above. In this case, the sensor and the vessel may be too far apart to transmit data to each other. An AUV with the \code{DtnLink} agent can help in relaying messages between these two vessels. The parameters for the same are given below in Table \ref{tab:datamuleparam} and Table \ref{tab:datamulenodes}.

% FIXME: get rid of this wall of text

\begin{table}[!h]
	\centering
	%	\resizebox{0.5\linewidth}{!}{%
	\begin{tabular}{|l|l|}
%		\toprule
		\hline
		\textbf{Parameter} & \textbf{Value} \\
		\hline
		Simulation Time & \SI{8800}{\second} \\
		Communication Range & \SI{600}{\meter} \\
		Message Size & \SI{50}{bytes} \\ 
		Message Frequency & \SI{10}{\second} \\
		Message TTL & \SI{8800}{\second} \\
		Total Messages sent from Source & 200 \\
		\hline
	\end{tabular}
	%	}
	\caption{Data Muling simulation parameters}
	\label{tab:datamuleparam}
\end{table}

\begin{table}[!h]
	\centering
	%	\resizebox{0.5\linewidth}{!}{%
	\begin{tabular}{|l	|c|c|c|}
		\hline
		\textbf{Node} & \textbf{X} & \textbf{Y} & \textbf{Z} \\
		\hline
		Sensor (Source) & \SI{0}{\metre} & \SI{0}{\metre} & \SI{-50}{\metre} \\
		AUV (Data Mule) & \SI{900}{\metre} & \SI{0}{\metre} & \SI{-50}{\metre} \\
		Ship (Destination) & \SI{1800}{\metre} & \SI{0}{\metre} & \SI{-50}{\metre} \\
		\hline
	\end{tabular}
	\caption{Initial node co-ordinates for the Data Muling scenario}
	\label{tab:datamulenodes}
\end{table}

In this particular simulation, the AUV starts near the sensor and makes it way to the research vessel. It makes two rounds in the trajectory shown in Figure \ref{fig:mulecomm}, with each round having a period of \SI{4400}{\second}. In this particular simulation, the research vessel and the sensor are kept apart at a distance of \SI{1800}{\metre}. The detection range of the nodes is set to \SI{600}{\metre}. The source node generates 100 messages with a TTL of \SI{8800}{\second} containing 50 bytes of randomly generated data which is sent out every \SI{10}{\second} for \SI{2000}{\second}. The entire simulation is run for \SI{8800}{\second} each for different values of \code{pDetection}.

\begin{figure}[h!]
	\centering
	\includegraphics[width=\linewidth]{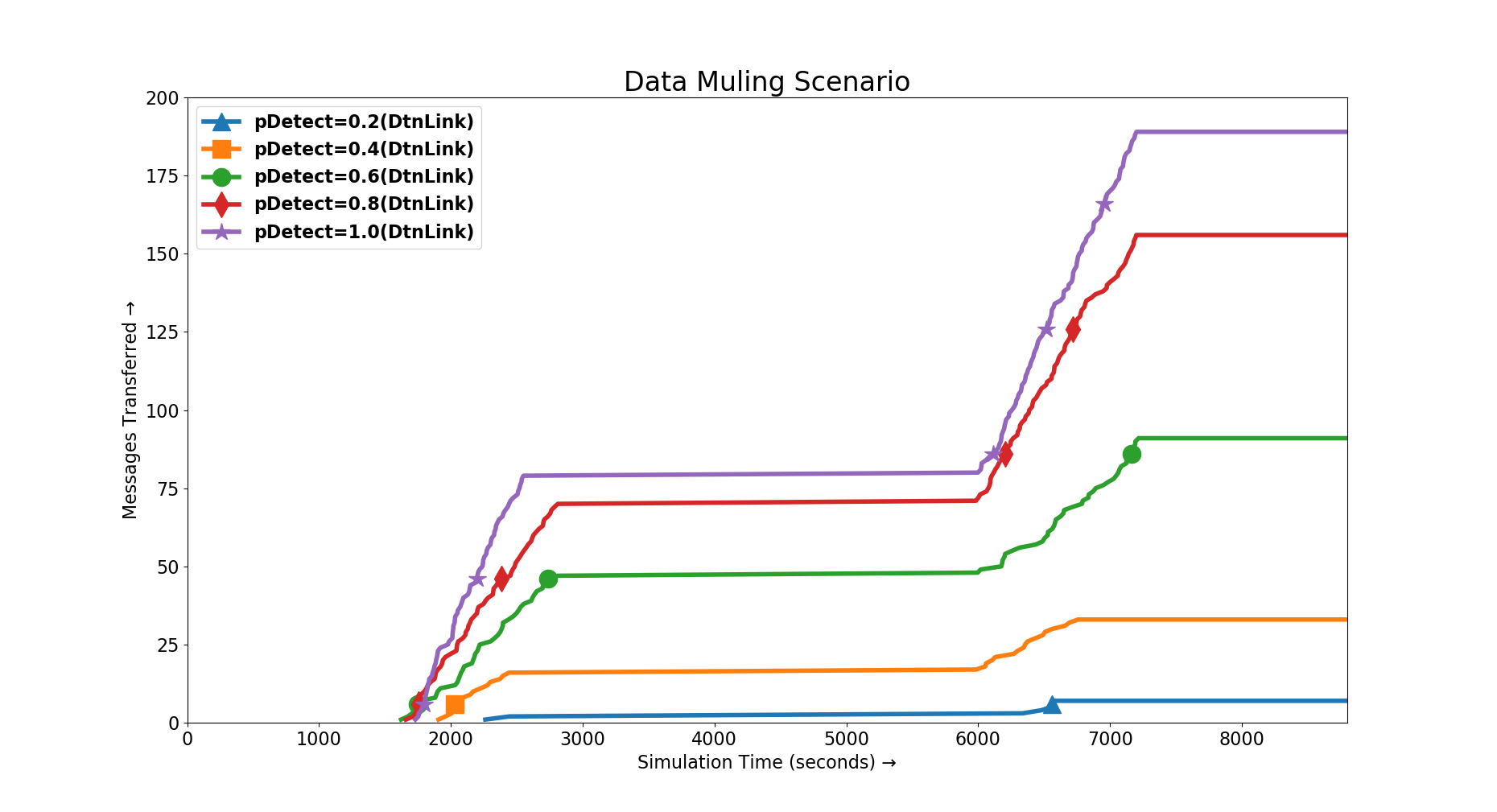}
	\caption{One Sender and one Receiver in the Data Muling scenario}
	\label{multiauv}
\end{figure}

Figure \ref{multiauv} illustrates this situation. At all values of \code{pDetection} we can see distinct trends in how messages are transferred. In the curve of messages delivered successfully at the destination, we can see that there are no messages transferred till \(t=\SI{1700}{\second}\), a spike in message delivery till \(t=\SI{2500}{\second}\), a period of dormancy till \(t=\SI{6000}{\second}\), and a final spike of message delivery which lasts till \(t=\SI{7200}{\second}\).

This can be explained by observing the periods of time in which the AUV comes in communication range of a node. At the beginning, the AUV collects data from the sensor while it is far away from the ship. This occurs till roughly \(t=\SI{900}{\second}\), after which the AUV moves away from the sensor, with its internal storage populated with messages to be transferred to the ship.

At around \(t=\SI{1700}{\second}\), the AUV passes by the ship and transfers its pending messages to it. This continues till the AUV has either moved out of the range of the ship or has finished sending whatever messages it picked up when it was in the range of the sensor earlier.

After this the AUV comes out of the communication range of the ship at around \(t=\SI{2500}{\second}\). It makes another flyby of the sensor and receives more messages that were pending on the sensor's internal storage. These are carried over and transferred to the ship at around \(t=\SI{6000}{\second}\) to transfer whatever new messages it received from the sensor between \(t=\SI{2500}{\second}\) and \(t=\SI{6000}{\second}\). The AUV then comes out of range of the ship at around \(t=\SI{7200}{\second}\) and returns to its starting point next to the sensor.

These trends in message delivery can be observed consistently at all values of \code{pDetection}. Using a \code{LINK} which does not support disruption tolerance would have caused all the messages meant for the ship in this scenario to fail instantly as the sensor and the ship are never in the communication range of the AUV at the same time. Therefore, we can see that \code{DtnLnk} can open up new options for network topologies which were not previously possible with other \code{LINK} agents.

%\begin{figure}[h!]
%	\centering
%	\includegraphics[width=\linewidth]{Figures/vlauvsim.png}
%	\caption{Both nodes send and receive}
%	\label{vlauv}
%\end{figure}
%
%Figure \ref{vlauv} shows a modified version of the above scenario. Here, both the nodes send and receive messages. Each node generates 100 messages and the simulation is allowed to run for a period of \SI{17600}{\second}. As each round of the AUV has a period of \SI{4400}{\second}, it can make a total of four rounds in this simulation. This figure correlates with this as there are four steps in the graph for each node where it receives messages from the AUV.

% Chapter 3

\chapter{Conclusions}

\label{Conclusions}

\lhead{\emph{Conclusions}}

Underwater communication is a rapidly developing field which is supplemented with acoustic communications, optical links, and AUVs. Due to various reasons such as interference due to the noise from ships, sounds from animals, and packet collisions in the channel medium, there are several challenges in successfully delivering a data underwater. UnetStack, the software stack of the Unet project allows one to deploy network protocols in software which can later be deployed on real hardware.

As shown in the scenario discussed in Section \ref{sec:dtn_hops}, \code{DtnLink} can significantly improve the success rate of message delivery without sending unnecessary transmissions, owing to the use of \code{Beacon} datagrams, timeouts for each link, and TTLs for each message. Furthermore, the data muling scenarios in Section \ref{sec:auv_mule} illustrate that \code{DtnLink} can open up new possibilities in network topologies which were not earlier possible with non-disruption tolerant \code{LINK} agents.

However, it is important to note that \code{DtnLink} may not be ideal in all use cases as demonstrated in Figure \ref{dvrl4} where it can be seen that \code{ReliableLink} is better than \code{DtnLink} in time-constrained applications. Hence, it is upto the discretion of the user to configure and use \code{DtnLink} with parameters (Section \ref{sec:config}) which best fit the environment where the protocol is to be deployed.

\code{DtnLink} uses an extensive \code{JUnit} test suite for each build for regression testing. Future work for the \code{DtnLink} includes expanding the concept to cover multi-hop acknowledgements with specialised underwater routing algorithms. It would also be useful to make \code{DtnLink} a smart protocol, which would be able to adapt to its environment according to its measurements of the channel's performance.

%\input{Chapters/Chapter4}
%\input{Chapters/Chapter5}
%\input{Chapters/Chapter6}
%\input{Chapters/Chapter7}

%-------------------------------------------------------------------------------
%	THESIS CONTENT - APPENDICES
%-------------------------------------------------------------------------------

\addtocontents{toc}{\vspace{0em}} % Add a gap in the Contents, for aesthetics

\appendix % Cue to tell LaTeX that the following 'chapters' are Appendices

% Include the appendices of the thesis as separate files from the Appendices
% folder
% Uncomment the lines as you write the Appendices

% Appendix A

\chapter{Appendix}

\label{Appendix} % For referencing this appendix elsewhere, use \ref{AppendixA}

\lhead{\emph{Appendix}} % This is for the header on each page - perhaps a shortened title
[\href{https://github.com/shortstheory/underwater-dtn/}{1}]
Source code for \code{DtnLink}

[\href{https://www.dropbox.com/s/wudkpl2wkpygkpx/rfc.pdf?dl=0}{2}]
RFC For Disruption Tolerant Protocols in UnetStack

[\href{https://www.dropbox.com/s/g0t11y2k8fltjkb/dtn-presentation.pdf?dl=0}{3}] Final thesis presentation

\addtocontents{toc}{\vspace{0em}} % Add a gap in the Contents, for aesthetics

\backmatter

%-------------------------------------------------------------------------------
%	BIBLIOGRAPHY
%-------------------------------------------------------------------------------

\label{Bibliography}

\lhead{\emph{Bibliography}} % Change the page header to say "Bibliography"

% Use the "unsrtnat" BibTeX style for formatting the Bibliography
\bibliographystyle{ieeetr}

% The references (bibliography) information are stored in the file named
% "Bibliography.bib"

\end{document}